\documentstyle[prl,aps,twocolumn,epsf]{revtex}
\draft
\begin{document}
\title{Fluctuation Induced Drift and Current Reversal
in Symmetric Potentials}
\author{Rangan Lahiri\cite{add1}}
\address{Department of Physics, 
Indian Institute of
Science,Bangalore 560 012, India.} 
\date{Version 12.7.96, printed \today }
\maketitle
\begin{abstract}
We explore the motion of a classical particle in a symmetric potential
with non-Gaussian skewed white  noise. 
We show analytically and numerically that the presence of nonzero
odd moments leads to a macroscopic current.
For a noise with a vanishing third moment we find that the current
changes sign as the potential or the noise strength is increased.
Possible physical situations are discussed including 
motor-protein motion and driven systems like fluidized beds. 
\end{abstract}


There can be no current in a system in thermal 
equilibrium. 
Drift is usually associated with the presence of an external
field, for example an electric field 
as in the case of charge drift.
However, it has been shown in recent studies 
\cite{proacadetc}
that a spatially asymmetric potential  
(usually taken to be an asymetric sawtooth)
which exerts  no macrosopic force,
combined with temporally periodic forcing or a noise of
non-thermal origin leads to non-equilibrium steady
states that carry current. 
It is believed that these models may have relevance 
in biological transduction and the motion of motor proteins in eukaryotic cells.

\begin{figure}
\epsfxsize=9cm 
\centerline{\epsfbox{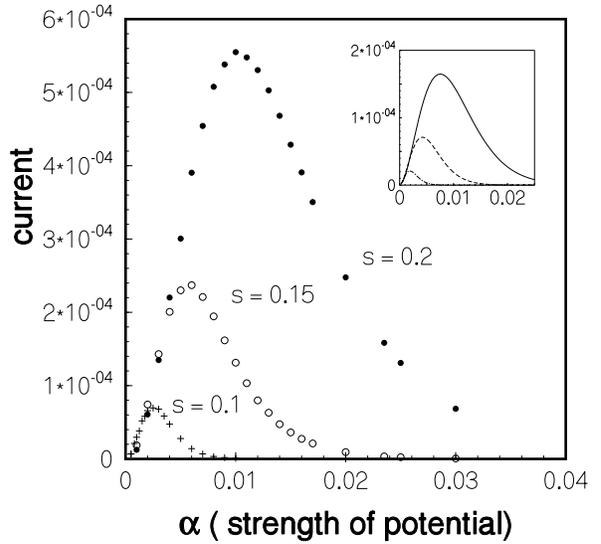}}
\caption{Variation of current with strength of confining 
potential: Monte Carlo Simulation in symmetric sawtooth potential.
The different curves correspond to different values of the noise strength.
The probablity of moving to the left and right respectively in
each step was $p_l = 0.2$ and $p_r = 0.1$.
The current plotted is the distance moved in one timestep,
where the period of the potential is $=1$.
The inset shows the current in the approximate analytic solution
of the master equation for the same set of parameters.}
\label{alphanum}
\end{figure}

\begin{figure}
\epsfxsize=9cm 
\centerline{\epsfbox{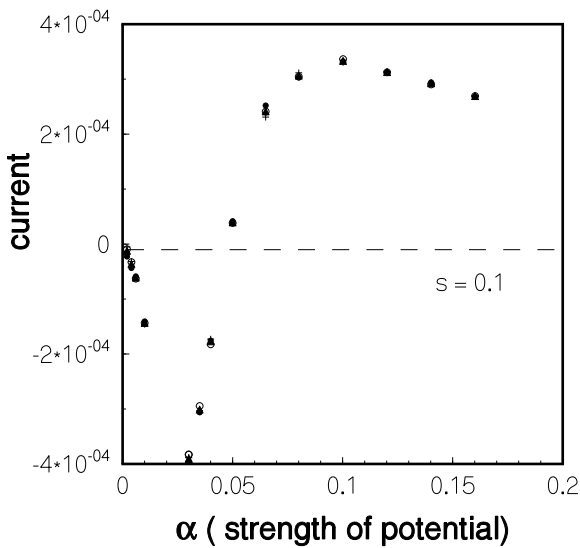}}
\caption{Reversal of the sign of current as potential strength is 
increased: Monte Carlo simulation on symmetric sawtooth potential
with a noise with three kicks such that first and third moments vanish.
The noise strength $s$ for this run was $0.1$
and the noise was composed of three kicks of strength $s$, 
$-2s$ and $3s$ with 
probabilities $\frac{1}{2} p_m$, $\frac{2}{5} p_m$ and 
$\frac{1}{10} p_m$, with $p_m = 0.3$.
The run was for $10^8$ steps.}
\label{reversalnum}
\end{figure}

In this paper we show that (1) it is not necessary to have a spatial
potential that violates parity to get unidirectional drift 
\cite{dae,other,luczka,millonassym,jayansym};
a combination of {\em symmetric} potential and non-thermal
skew noise is sufficient;
(2) the current vanishes in the limits of very strong or very weak
potentials or noise amplitudes (Fig. \ref{alphanum});
(3) if the skew noise is dominated by a third moment then the
current always has the same sign as the third moment(Fig.
\ref{alphanum}, inset);
(4) higher moments can lead to current reversal (Fig. \ref{reversalnum}).
Finally we  construct a Maxwell demon based on our model and also
suggest an experiment involving a colloidal particle in a fluidized bed.

Perhaps the simplest physical example of a skew noise occurs in the
case of a Brownian particle falling under gravity in an ideal gas. 
The noise is the force in the rest frame of the particle 
due to random kicks coming from collisions with gas molecules. 
Since the net force in the rest frame is zero, the kicks 
in the downward direction, though fewer in number, have to be stronger, 
leading to a skew noise.
Besides this specific example, 
the noise is expected to be skewed in
any driven system where the driving violates parity.
e.g. colloidal suspensions in a fluidized bed, 
driven diffusive lattice gases etc.  

To capture the essentials of the problem,
we consider overdamped motion of a particle in a periodic
symmetric sawtooth potential  $V(x)$ with a skewed noise.
Suppose the noise kicks come at regular intervals of time $\Delta$.
The displacement in one such interval, $\delta x = \alpha +\eta,$
has a deterministic part
$\alpha = - \Gamma \frac{\partial V}{\partial x} \Delta $ 
where $\Gamma $ is the inverse of the friction coefficient,
and a stochastic part $\eta$, defined by:
\vspace{0.25cm}\\
\begin{tabular}{p{2.5cm}p{2.5cm}p{3cm}}
$\eta = -s/3 $ & with probability &$p_l = \frac{2}{3} p_m$\\ 
$\eta = +2s/3 $ & with probability &$p_r = \frac{1}{3} p_m$\\ 
$\eta = 0 $ & with probability &$1-p_m$
\end{tabular}
The noise is uncorrelated in time by construction.
$p_0 = (1-p_m) $ is the probability that the particle does not 
experience a noise kick during the interval.

If the stochastic part of the displacement is much smaller
than the  the length $l_V$ over which the deterministic 
force changes appreciably ($l_V =1$ in our case), then it is legitimate
to consider a larger timestep $N \Delta$, within which $N$ noise
kicks are present, adding up to a Gaussian thermal distribution
which can cause no drift.
If, however, the displacement due to noise is comparable to
$l_V$, the local environment changes considerably after each
noise-kick, so that it is no longer permissible to add several
realizations and the non-Gaussian nature of the noise manifests itself. 
In the extreme limit where the potential is weak
and the noise strength $s$ lies between
$\frac{1}{2}$ and $1$ (where the potential has a period 1),
it is easy to see qualitatively that there will be a drift 
in the direction of the stronger but less frequent kicks.
It turns out that even
for the less obvious case of weaker noise amplitudes, there
is a current towards the right, which goes continuously to zero as the
noise amplitude is decreased.

Direct Monte Carlo simulations have been carried out on this model
(Fig. \ref{alphanum}). 
The simulation was run for a long enough time ($10^8$ steps)
so that the drift (which goes linearly in time) was significant compared 
to the free diffusion (i.e., the sum of the random numbers), which goes as
$\sqrt{t}$, and the current had acquired a steady value.
The drift current is expected to go to zero in two limiting cases
for different reasons. When the potential is
weak or the noise strength is large, the system
behaves as though there were no potential. On the other hand, 
when the potential is strong or the noise is weak, the particle is confined. 
The current is always positive, i.e., along  the direction of the
stronger kicks. 
This, as we shall see below, is a result of the third moment of the
noise playing the dominant role.

We now look at the evolution of the probability density in
the above problem.
First consider a general Markoff process, one in which the random
noise can have moments higher than the second. 
Let $P(x,n)dx$ be the probability of finding the particle between
$x$ and $x + dx$ at timestep $n$.
The time evolution of 
$P(x,n)$ is given by $P(x,n+1) = \int du P(x-u,n) T_u (x-u),$
where $ T_u (x)$ is the probability of transition from $x$ to
$x+u$ in one timestep.
The functions inside the integral on the right hand side 
can be Taylor expanded about $x$ to give 
an evolution equation for $P$ involving 
the moments of $x$ \cite{chandra}.
The rate of change of probability is the divergence of a current, given by
$P(x,n+1) - P(x,n) = - \frac{\partial}{\partial x} J$,
where the current is 
\begin{eqnarray}
J = 	             \left< u \right> P(x,n)
	            - \frac{1}{2!}\frac{\partial}{\partial x}
		    [ \left< u^2 \right> P(x,n)] \nonumber \\
	            + \frac{1}{3!}\frac{\partial^2}{\partial x^2 }
		    [ \left< u ^3\right> P(x,n)] ...
\label{current}
\end{eqnarray}
The moments 
$\left< u^m \right>(x) = \int du u^m T_u (x) $
that enter into this equation are the combined effect
of the potential and the noise. For the model described above, where 
the displacement in timestep $\Delta$ is 
$\alpha(x) = - \frac{\partial V}{\partial x}\Delta;$ due to the
potential and $-s_l$ and $s_r$ with probabilities 
$p_l$ and $p_r$ respectively due to the noise,
the moments are
$\left< u \right> = \alpha;$
$\left< u^2 \right> = \alpha^2 + \mu_2;$
$\left< u^3 \right> = \alpha^3 + 3 \alpha \mu_2 + \mu_3;$
and so on,
where $\mu_n$ is the $n^{\mbox{th}}$ moment of the noise:
$\mu_n=p_l (-s_l)^n + p_r s_r^n$.
For an arbitrary potential, $\alpha$ is a function of x whereas
the $\mu_n$s are constants.

For ordinary Brownian motion, the series in equation (\ref{current})
may be truncated at the second term.
To see that there is no current in this case, 
simply divide equation \ref{current} (truncated at the second order)
by $P$ and integrate both sides 
from $0$ to $1$, with the limits identified with each other due to 
periodic boundary conditions.
The first moment is caused by the static potential alone,
since noise has a zero mean.
It is antisymmetric about $\frac{1}{2}$ for a symmetric potential,
so its integral is zero.
The second term is a total derivative, 
therefore it's integral also vanishes on a circle.
Since the integral of $\frac{1}{P} $ on the left hand side is
positive definite, $J$ must be zero.

Let us now restrict ourselves to processes in which the $n^{th}$
moment is given by $\left< u^n \right> = \mu_n.$
This amounts to assuming that $\alpha / s $ is small, so that
the entire contribution to the second and higher moments 
comes from the noise alone, 
an approximation which holds good for sufficiently weak potentials.
The current can can again be calculated as above, but
the additional terms on the RHS are no longer 
total derivatives.
We also assume that the higher moments ${\mu}_n$ of the noise 
are negligible compared to the third moment ${\mu}_3$.
Under these approximations the current turns out to be
\begin{equation}
J  = 
                    \frac{1}{3!}
                    \mu_3
                    \frac{ \int dx
		    \frac{1}{P}  
                    (\frac{\partial  P}{\partial x })^2}
	       	    {\int \frac{dx}{ P }},
\label{currentfor3rdmoment3}
\end{equation}
which clearly has the same sign as the third moment $\mu_3$
in agreement with intuitive expectations as well as our numerical results.

When the third moment is much smaller than the second, i.e., for
slight deviations from a thermal distrbution, we can calculate this current
perturbatively. The probability distribution is the equilibrium distribution
plus a correction due to the third moment,
$P(x) \sim e ^{- \beta V(x)} + o(\mu_3).$
Since the current is already first order in $\mu_3$, we can use the 
equilibrium probability distribution itself in equation 
(\ref{currentfor3rdmoment3})
to get
\begin{equation}
     J =            \frac{1}{3!}
                    \mu_3 {\beta}^2
                    \frac{ \int dx
                    \left(\frac{\partial V}{\partial x } \right)^2}
	       	    {\int dx e^{\beta V(x)}}
		    + o({\mu _3}^2).
\label{currentfor3rdpert}
\end{equation}
As expected, this current goes to zero when the potential becomes flat.
Curiously, if we add a slight disorder to the potential, 
thereby making it more corrugated, i.e.,
give it a larger value of $({\frac{\partial V}{\partial x})}^2$
we get an {\em enhancement} in the current.

To express the current (\ref{currentfor3rdpert})
in terms of the parameters $\alpha$ and $s$, 
we compare the small-kick limit of the discrete model with 
the equation of motion of a particle in a potential $V$
and temperature $T$.
This gives us the connection:
$\alpha |x| = \Gamma \Delta V(x)$
for the potential and 
$2 \Gamma k_B T \Delta =  \left(\frac{ p_l p_r s^2}{p_l+p_r} \right)$ 
for the noise.
Equation (\ref{currentfor3rdpert})  then gives
$ J = \frac{4}{3!} \mu_3 {\lambda ^3}/({e^{\lambda} -1})$
where
$\lambda = 
\frac{\alpha (p_r + p_l)}{p_r p_l s^2}.$
This current is shown in Fig. \ref{alphanum} (inset). 
In spite of the drastic approximations 
there is excellent qualitative agreement with the numerical result.
As expected, the drift vanishes in the limiting cases of 
(i) weak potentials or strong noise and
(ii) strong potentials or weak noise, 
in a manner which we now calculate.

For $\lambda \longrightarrow 0$ and we have
$J \sim \mu_3 
{\left( \frac{\alpha}{s^2} \right)}^2$
showing that $J \sim \alpha ^2$ in the limit of weak potential strength.
Since, $\mu _3 \sim s^3$
the current falls off as $s^{-1}$ as the 
noise strength $s$ is made large.
On the other hand, for large $\lambda$,
$J \sim \frac{1}{s^3} exp\left[- \frac{\mbox{constant}}{s^2} \right].$
This is the manner in which the current approaches zero as the noise
kicks $s$ become smaller and the validity of the central limit theorem 
improves.

Interesting features develop when
the noise is dominated by the fifth moment. 
If we neglect all other moments, The current
can again be calculated perturbatively:
\begin{equation}
     J =            \frac{1}{5!}
                    \mu_5 
                    \frac{ \int dx \left[
      {\beta}^4 \left(\frac{\partial  V}{\partial x } \right)^4
     -  {\beta}^2 \left(\frac{\partial ^2 V}{\partial x^2 } \right)^2
			\right]}
	       	    {\int dx e^{\beta V(x)}}.
\label{currentfor5thpert}
\end{equation}
The second term would lead to a current {\em opposite} in sign 
with respect to
the fifth moment. For a sawtooth potential, the integral of
$\left(\frac{\partial ^2 V}{\partial x^2 } \right)^2$
is ill defined. If, however we replace the sawtooth by a potential
with a rounded wedge so that the derivative 
$\frac{\partial V}{\partial x}$
changes from $-\alpha$ to $+\alpha$
over a length $\epsilon$, then the current is
\begin{equation}
     J =            \frac{16}{5!} \mu_5 
		\frac
                {\lambda ^5 - 2 \frac{\lambda ^3}{\epsilon}}
		{e^{\lambda} -1}.
\label{j5}
\end{equation}
Evidently, the negative part of the current dominates for small
$\lambda $, i.e., for weak potentials $\alpha$ or large noise strength $s$.
The sign changes as the strength $\alpha$ is increased.
We have found such a reversal in direct simulations as well
( Fig. \ref{reversalnum} ).
Current reversal has been found in the literature 
\cite{reversals}
for a particle in an ratchetlike potential with coloured 
noise.
In the present model, the presence of higher moments in a white
noise causes current reversal as a parameter, e.g.,
the potential strength is changed continuously.

Having explored the consequences of the model in some detail, we
now go back to physical realizations.
Before going on to a concrete experimental situation where skew noise
may arise, let us perform a thought experiment.
Consider two `gases' having different particle masses kept in 
separate containers, say Hydrogen (lighter
atoms) and Nitrogen (heavier). Let both the containers be at the 
same temperature and pressure, and hence the same density (the
densities are sufficiently low so that the gases can be treated as
ideal). Now connect the reservoirs by a long tube which has
a partition in the middle separating the two. 
The pressure is the same on both sides so the partition remains stationary.
Note, however, that the random force that the partition experiences 
is a noise precisely of the kind discussed above \cite{fndemon}. 
Now, if there are minute corrugations on the base of the tube then
they supply a periodic potential due to the gravitational field, and
the partition should start moving into the lighter gas.
It is possible even to tilt the tube slightly and drive the partition
up against the gravitational field, thereby extracting work from a single
system at thermal equilibrium!

The demon can be rationalized along lines similar to other
more familiar ones\cite{feynman}. 
We have assumed overdamped dynamics of the partition and
neglected the noise associated with it, which is always there in any
source of dissipation.
If the corrugated floor is colder than the reservoirs, then this noise
amplitude ( which, for a given damping, is proportional to the square
root of temperature) is negligible, and the partition drifts.
In this situation we are driving an engine between the gases at a higher
temperature and the floor at a lower temperature.
As the partition moves, it keeps transferring heat to the floor,
until it reaches the temperature of the reservoirs.
Now the thermal fluctuations of the partition completely wash out the
effect of the asymmetric noise (coming from collisions) and the drift stops.

A more concrete experimental situation may be realized
using charged colloidal particles in a fluidized bed. 
A steady upward flow of fluid supplies the Stokes friction that is
adjusted to balance exactly the weight of the particle.
Brownian motion is negligible for most experimental systems, but
there are indications of the existence of a `noise' on the particle
due to hydrodynamic interactions\cite{bossis}.
The source of this noise is completely athermal.
From symmetry considerations, this noise is expected to be 
skewed, in a way similar to the noise on a falling particle
in an ideal gas.
If we now switch on a sinusoidal laser field forming standing waves
with an intensity variation in the vertical direction,
the particles will start drifting downwards in the lab frame.
The drift speed will depend on the flow rate in the fluidized bed
(which controls the noise amplitude $s$)
and the amplitude of the laser field ($\sim \alpha$). 

Finally, a word about biology. The `standard model'
with a ratchet-like potential and some kind of correlated
noise has been advertised as a candidate for 
modelling motor protein motion in eukaryotic cells. 
However, there appears to be no direct evidence of such ratchet potentials
in actual experiments.
The model presented here can be an equally well suited
candidate for motion of proteins and has
all the essential features of the standard model,
including a possible reversal of current as a parameter is 
continuously changed.
The question of current reversal is interesting because
motor proteins of the same class are known to move along
different directions; 
for example a drosophilla kinesin, Ncd, and axonal kinesin
walk in opposite senses along a microtubule \cite{watson}.
Our model gives, for the case where the fifth moment dominates,
a reversal of current simply by changing the noise 
amplitude (which should correspond to ATP activity)
leading to the possibility of two-way traffic along the highways.

In conclusion, we have shown that it is possible to produce a
macroscopic current in a parity non-violating periodic potential
if the noise has nonzero odd moments.
The current vanishes when either the potential strength
or the noise amplitude become very small or very large
and exhibits sign reversal if the third moment is negligible.
The vanishing of the current in the strong noise limit
distinguishes our model from that of Luczka et. al. \cite{luczka}
who consider a symmetric potential and a shot noise 
and find a saturation of the current in the strong-noise limit.
This is because of the extreme nature of the shot noise such as the 
divergence of moments higher than the second.

I thank Pinaki Majumdar for drawing my attention to this exciting
class of problems, Joseph Samuel for suggesting
the Maxwell demon,
Jayanth Banavar for a critical reading of the manuscript
and Arun Jayannavar for pointing out references \cite{luczka} and 
\cite{millonassym} after completion of this work.
I have benefitted from discussions with
Abhishek Dhar,
Anirban Sain, Arun Jayannavar, Arvind, H. R. Krishnamurthy, 
N. Kumar, Pinaki Majumdar, 
Jacques Prost, Rahul Siddharthan, Rajaram Nityananda, Sriram Ramaswamy, 
Supurna Sinha, Sumit Banerjee, V. Suri and Sushan Konar. 
Financial support from CSIR, India and computational facilities
provided by SERC, IISc are duly acknowledged.


\begin{references}
\bibitem{other}
This result has been obtained independently by other authors 
\cite{luczka,millonassym,jayansym}
for different physical situations. We compare the results at the end
of the paper.
\bibitem{fndemon}
The momentum imparted to the partition in a collision is $2 p$
where $p = \sqrt{2 m k_B T} $. For the heavier particles this is larger
but is compensated by a smaller number of collisions, which is
proportional to $v = \sqrt{2 k_B T / m}$.
\end{references}
\end{document}